\documentclass[12pt]{amsart}

\usepackage[T1]{fontenc}  
\usepackage[utf8]{inputenc}  
\usepackage{lmodern}  
\usepackage{amsmath, amssymb, amsfonts, amscd}  

\usepackage{graphicx}  
\usepackage{xcolor}  
\usepackage{subcaption}  
\usepackage[all]{xy}  

\usepackage{geometry}
\theoremstyle{definition}

\theoremstyle{remark}

\usepackage{hyperref}  
\hypersetup{
    colorlinks=true,
    linkcolor=blue,
    citecolor=blue,
    urlcolor=blue,
    pdfauthor={Your Name},
    pdftitle={Your Paper Title}
}

\begin{document}
\title{Causality and thermodynamics in anisotropic 
fluids}

\author{Jos\'e Díaz Polanco}
\address{Instituto de Ciencias Exactas y Naturales, Facultad de  Ciencias, Universidad Arturo, Prat Iquique, Chile}
\email{joseludi@unap.cl}

\author{Jos\'{e} Ayala Hoffmann}
\address{Universidad de Tarapac\'a, Iquique, Chile}
\email{jayalhoff@gmail.com}

\author{Luis Viza}
\address{Facultad de  Ciencias, Universidad Arturo, Prat Iquique, Chile}
\email{luisvizagarcia@gmail.com}
\maketitle

\begin{abstract} \baselineskip=1.2\normalbaselineskip
We propose a thermodynamic formalism, within the particle-frame, for the energy-momentum tensor of irreversible anisotropic imperfect fluids subject to causality. Building on the Israel-Stewart extension of Eckart's theory, we further generalize these formalisms to incorporate anisotropic effects while ensuring the preservation of causality. In this framework, the second law of thermodynamics includes an additional term accounting for the system's anisotropy, which we derive explicitly in closed form for both first- and second-order theories. Notably, when anisotropy is removed, our model recovers Eckart's theory at first order and Israel-Stewart's at second order.
\end{abstract}

\section{Introduction} \baselineskip=1.2\normalbaselineskip

Astrophysical systems are often modeled as rotating fluids influenced by various physical fields and local phenomena, as for example, strong gravitational forces, electromagnetic fields, high temperature gradients, nuclear reactions, and phase transitions. Under such extreme conditions, anisotropic and dissipative behaviours emerge in contrast to the assumptions of isotropy and equilibrium that underlie traditional Newtonian and relativistic models \cite{{oppenheimer1939},tolman1939}.

In the relativistic framework, a sufficiently general theoretical model is essential, not only to describe the complex dynamics and stability of these fluid systems but also to address the fundamental issue of causality inherent in Eckart’s first-order relativistic thermodynamics \cite{eck}.

A suitable framework is given by the Israel-Stewart theory \cite{is1, is2}, which introduces a second-order thermodynamic formulation that respects causality \cite{his}. This approach considers an arbitrary state of an isotropic imperfect fluid described by a symmetric and conserved energy-momentum tensor, a conserved particle flux vector, and an entropy flux satisfying the laws of the thermodynamics.

In practice, this energy-momentum tensor appears on the right-hand side of Einstein’s equations, which encapsulates all physical interactions within the system. Typically, multiple energy-momentum tensors must be combined, requiring careful thermodynamic treatment \cite{DAU}. In this regard, it is important to remember that the coupling of two energy-moment tensors like perfect fluids can be effectively described as an anisotropic fluid \cite{letelier1980}. This suggests that anisotropy naturally emerges from the coupling of two or more fluid components, each described by its own energy-momentum tensor, within a consistent thermodynamic framework. 

We study causality and thermodynamics in anisotropic fluids using a particle-frame description within a consistent thermodynamic framework. The anisotropy in this formulation is characterized by a single parameter, which represents a fraction of the effective viscous pressure and determines the viscous pressure in each spatial direction. It is noteworthy that, in the absence of anisotropy, our formalism reduces exactly to Eckart's first-order theory and to Israel-Stewart's second-order theory. 

The main contributions of this article are as follows.

\begin{enumerate}

\item We extend the first- and second-order relativistic thermodynamic formalism to anisotropic fluids, ensuring that our model precisely recovers Eckart's theory at first order and Israel-Stewart's theory at second order in the isotropic limit.

\item We explicitly derive all the equations associated with the components of the energy-momentum tensor for anisotropic imperfect fluids, ensuring consistency with thermodynamic principles and causality.

\item The variation in the entropy flux equation, which is proven to be increasing, introduces an additional term directly related to the viscous pressure along the anisotropic direction. We explicitly derive this term for both, first- and second-order formalisms.
\end{enumerate}


This paper is structured as follows. In Section \ref{two}, we present an overview of the energy-momentum tensor for anisotropic fluids, introducing notation and fundamental concepts used throughout the paper. In Section \ref{three}, we study anisotropy constrained to a single spatial direction, discussing its implications within the relativistic thermodynamics framework. Section \ref{four} develops a first-order thermodynamic formalism for anisotropic fluids, explicitly deriving the associated equations and ensuring thermodynamic consistency. Extending this analysis, Section \ref{five} introduces a second-order thermodynamic framework, incorporating causality and stability conditions based on the Israel-Stewart formalism. Finally, Section \ref{six} summarizes the main findings and contributions of this work, highlighting potential applications and suggesting directions for future research.



\section{Overview of the Energy-Momentum Tensor for Anisotropic Fluids}\label{two} 

Without loss of generality, we adopt a tetrad basis $\widehat{e}{(j)}$ 
aligned with the diagonal metric $g_{\alpha \beta}$, ensuring the standard decomposition:
\begin{equation}
g^{\alpha \beta }=\eta ^{(j)(k)}\widehat{e}_{(j)}^{\alpha }\widehat{e}%
_{(k)}^{\beta }=-\widehat{e}_{(0)}^{\alpha }\widehat{e}_{(0)}^{\beta }+%
\widehat{e}_{(1)}^{\alpha }\widehat{e}_{(1)}^{\beta }+\widehat{e}%
_{(2)}^{\alpha }\widehat{e}_{(2)}^{\beta }+\widehat{e}_{(3)}^{\alpha }%
\widehat{e}_{(3)}^{\beta }  \label{eq:11}
\end{equation}%

 where $\eta_{(j)(k)}$ represents the Minkowski metric $(-1, 1, 1, 1)$, and repeated indices imply the sum from $0$ to $3$ following Einstein's convention. As usual, latin indices correspond to the tetrad, while the Greek indices refer to the respective components of the tetrad. The inverse relation of (\ref{eq:11}) follows immediately as
$g_{\alpha \beta }\widehat{e}_{(j)}^{\alpha }\widehat{e}_{(k)}^{\beta}=\eta
_{(j)(k)}.  \label{eq:12}$

In this way, the energy-momentum tensor in a generic tetrad basis can be expressed as:
 
\begin{equation}
T^{\alpha \beta }=a^{jk}\widehat{e}%
_{(j)}^{\alpha }\widehat{e}_{(k)}^{\beta}  \label{eq:13}
\end{equation}

where $a^{jk}$ are the components of the tensor to be determined, so that they represent a generic imperfect anisotropic fluid. The symmetry condition of the tensor $T^{\alpha \beta}$ implies that $a^{jk} = a^{kj}$. Thus, the energy-momentum tensor can be expressed in expanded form as:

\begin{eqnarray}
T^{\alpha \beta } &=&a^{00}\widehat{e}_{(0)}^{\alpha }\widehat{e}%
_{(0)}^{\beta }+a^{11}\widehat{e}_{(1)}^{\alpha }\widehat{e}_{(1)}^{\beta
}+a^{22}\widehat{e}_{(2)}^{\alpha }\widehat{e}_{(2)}^{\beta }+a^{33}\widehat{%
e}_{(3)}^{\alpha }\widehat{e}_{(3)}^{\beta }  \notag \\
&+&a^{10}\left( \widehat{e}_{(0)}^{\alpha }\widehat{e}_{(1)}^{\beta }+%
\widehat{e}_{(1)}^{\alpha }\widehat{e}_{(0)}^{\beta }\right) +a^{20}\left( 
\widehat{e}_{(0)}^{\alpha }\widehat{e}_{(2)}^{\beta }+\widehat{e}%
_{(2)}^{\alpha }\widehat{e}_{(0)}^{\beta }\right) +a^{30}\left( \widehat{e}%
_{(0)}^{\alpha }\widehat{e}_{(3)}^{\beta }+\widehat{e}_{(3)}^{\alpha }%
\widehat{e}_{(0)}^{\beta }\right)   \notag \\
&+&a^{21}\left( \widehat{e}_{(1)}^{\alpha }\widehat{e}_{(2)}^{\beta }+%
\widehat{e}_{(2)}^{\alpha }\widehat{e}_{(1)}^{\beta }\right) +a^{31}\left( 
\widehat{e}_{(1)}^{\alpha }\widehat{e}_{(3)}^{\beta }+\widehat{e}%
_{(3)}^{\alpha }\widehat{e}_{(1)}^{\beta }\right) +a^{32}\left( \widehat{e}%
_{(2)}^{\alpha }\widehat{e}_{(3)}^{\beta }+\widehat{e}_{(3)}^{\alpha }%
\widehat{e}_{(2)}^{\beta }\right)   \label{eq:14}
\end{eqnarray}%

The 4-velocity of the fluid as comoving is $u^{\alpha }=\widehat{e}_{(0)}^{\alpha } \rightarrow ( u^{0},0,0,0 )$. Additionally, according to Eq. (\ref{eq:13}), we have that $g_{\alpha \beta }\widehat{e}_{(0)}^{\alpha }\widehat{e}_{(0)}^{\beta }=\eta
_{(0)(0)}$ being this timelike and equivalent to $g_{\alpha \beta }u^{\alpha
}u^{\beta }=-1$. With this observation Eq. (\ref{eq:11}) can be written in the form:

\begin{equation}
g^{\alpha \beta }+u^{\alpha }u^{\beta }=\widehat{e}_{(1)}^{\alpha }\widehat{e%
}_{(1)}^{\beta }+\widehat{e}_{(2)}^{\alpha }\widehat{e}_{(2)}^{\beta }+%
\widehat{e}_{(3)}^{\alpha }\widehat{e}_{(3)}^{\beta },  \label{eq:16}
\end{equation}%

which defines a tensor with the property of being orthogonal to the 4-velocity corresponding to a spatial projector given by $h^{\alpha \beta} = g^{\alpha \beta} + u^{\alpha}u^{\beta}$. On the other hand, from \cite{eck} we know that the energy density $\xi$ and the effective mechanical pressure of the system $P_{ef}$ can be computed as follows:
 
\begin{equation}
\xi =T^{\alpha \beta }u_{\alpha }u_{\beta }\hspace{2cm} P_{ef}=\frac{1}{3}%
T^{\alpha \beta }h_{\alpha \beta } .\label{eq:17}
\end{equation}

The effective pressure reveals that the components $a^{11}$, $a^{22}$, and $a^{33}$ correspond to the pressures in the respective spatial directions, such that $a^{11} = P_{1}$, $a^{22} = P_{2}$, and $a^{33} = P_{3}$.  Furthermore, the energy-momentum tensor (\ref{eq:14}) can be substantially simplified if we define the 4-vector $Q^{\alpha}$ (heat flux) and the stress tensor $\Omega^{\alpha \beta}$  as follows,
\begin{equation}
Q^{\alpha} = a^{10}\widehat{e}_{(1)}^{\alpha} + a^{20}\widehat{e}_{(2)}^{\alpha} + a^{30}\widehat{e}_{(3)}^{\alpha},
\label{eq:20}
\end{equation}

\begin{equation}
\Omega ^{\alpha \beta }=a^{21}\left( \widehat{e}_{(1)}^{\alpha }\widehat{e}%
_{(2)}^{\beta }+\widehat{e}_{(2)}^{\alpha }\widehat{e}_{(1)}^{\beta }\right)
+a^{31}\left( \widehat{e}_{(1)}^{\alpha }\widehat{e}_{(3)}^{\beta }+\widehat{%
e}_{(3)}^{\alpha }\widehat{e}_{(1)}^{\beta }\right) +a^{32}\left( \widehat{e}%
_{(2)}^{\alpha }\widehat{e}_{(3)}^{\beta }+\widehat{e}_{(3)}^{\alpha }%
\widehat{e}_{(2)}^{\beta }\right),
\end{equation}%
note that the stress is a trace-free symmetric tensor.

Finally, the form of the energy-momentum tensor for a generic anisotropic fluid is given by
\begin{equation}
T^{\alpha \beta }=\xi u^{\alpha }u^{\beta }+P_{1}\widehat{e}_{(1)}^{\alpha }%
\widehat{e}_{(1)}^{\beta }+P_{2}\widehat{e}_{(2)}^{\alpha }\widehat{e}%
_{(2)}^{\beta }+P_{3}\widehat{e}_{(3)}^{\alpha }\widehat{e}_{(1)}^{\beta
}+Q^{\alpha }u^{\beta }+Q^{\beta }u^{\alpha }+\Omega ^{\alpha \beta }.
\label{eq:21a}
\end{equation}%

Note that this tensor is anisotropic in all directions. And, it is easy to verify that \( Q^{\alpha }u_{\alpha }=0 \) and \( \Omega ^{\alpha \beta }u_{\alpha }=0 \). In fact, we can show that
\begin{equation}
Q^{\alpha }=h_{\beta }^{\alpha }T^{\beta \gamma }u_{\gamma } \label{eq:21}
\end{equation}
and
\begin{equation}
\Omega ^{\alpha \beta }=h_{\gamma }^{\alpha }h_{\lambda }^{\beta }T^{\gamma \lambda } \label{eq:22}
\end{equation}
as in the standard theory of dissipative fluids \cite{eck}.

\section{Anisotropy Constrained to a Single Spatial Direction} \label{three}

Anisotropy in a single direction is a key feature of astrophysical phenomena driven by extreme conditions. It often aligns with magnetic field lines, influencing the behavior of plasma, heat flow, and electromagnetic radiation \cite{krause1980}. This anisotropy plays a critical role in understanding processes such as black hole jets \cite{blandford1982} and the dynamics of planetary rings \cite{goldreich1982}. In this section, we examine the appropriate covariant thermodynamics for an anisotropic fluid with a single preferential direction.

To explore such anisotropic effects on thermodynamic variables, we define a generic preferential direction, denoted \( x^{3} \). From now on, to maintain consistency with the astrophysical literature, we will use spherical coordinates, represented as \( (x^{1}, x^{2}, x^{3}) = (r, \theta, \phi) \).

In this framework, the components of the effective pressure are specified as \( a^{11} = a^{22} = P_{r} \) and \( a^{33} = P_{\varphi} \), where \( P_{r} \) and \( P_{\varphi} \) represent orthogonal pressures near a fictitious local equilibrium state \cite{roy}. In this state, the 4-velocity is chosen so that the energy density \( \xi \) matches that of the local equilibrium. It is crucial to emphasise that local equilibrium states vary between distinct events, further highlighting the importance of anisotropy in capturing the dynamics of astrophysical systems.

Under these all these conditions, the energy-momentum tensor in (\ref{eq:21a}) can be written as:

\begin{equation}
T^{\alpha \beta }=\xi u^{\alpha }u^{\beta }+P_{r}h^{\alpha \beta }+\left(
P_{\varphi }-P_{r}\right) \widehat{e}_{(3)}^{\alpha }\widehat{e}%
_{(3)}^{\beta }+Q^{\alpha }u^{\beta }+Q^{\beta }u^{\alpha }+\Omega ^{\alpha
\beta },  \label{eq:23}
\end{equation}

where $h^{\alpha \beta }$ is the spatial projector, $Q^{\alpha }$ is the heat flux, and $\Omega^{\alpha \beta }$ the stress tensor as defined in Section \ref{two}. 

Now we contract the energy-momentum tensor in Eq. (\ref{eq:23}) with the 4-velocity comoving with the fluid defined as $u^{\alpha }=\widehat{e}_{(0)}^{\alpha }$, in the Section \ref{two} to obtain, 
\begin{equation}
u_{\beta }T^{\alpha \beta }=-\xi u^{\alpha }-Q^{\alpha }
\label{eq:24}
\end{equation}

which represents the conservation of energy fluxes within the system. We proceed by applying the covariant derivative, leveraging the condition $%
\nabla_{\beta} T^{\alpha \beta }=0$ to derive
\begin{equation}
T^{\alpha \beta }\nabla _{\alpha }u_{\beta }=-u^{\alpha }\nabla _{\alpha
}\xi -\xi \nabla _{\alpha }u^{\alpha }-\nabla _{\alpha }Q^{\alpha },
\label{eq:25}
\end{equation}
which, after using Eq. (\ref{eq:23}), takes the form
\begin{equation}
P_{r}\theta +\left( P_{\varphi }-P_{r}\right) \digamma+Q^{\beta }u^{\alpha
}\nabla _{\alpha }u_{\beta }+\Omega ^{\alpha \beta }\nabla _{\alpha
}u_{\beta }=-u^{\alpha }\nabla _{\alpha }\xi -\xi \nabla _{\alpha }u^{\alpha
}-\nabla _{\alpha }Q^{\alpha }  \label{eq:26}
\end{equation}
where $\theta=h^{\alpha \beta} \nabla _{\alpha }u_{\beta }$ is known as the expansion, and $\digamma =\widehat{e}_{(3)}^{\alpha }\widehat{e}_{(3)}^{\beta }\nabla
_{\alpha }u_{\beta }$ 
is a scalar associated with the dissipation in the preferred direction, whose physical interpretation will be clarified subsequently. 

From now on, we will refer to the effective pressure $P_{{ef}}$ as the equivalent mechanical pressure for the anisotropic fluid, denoted by $P_{m}$. In addition, by using Eq. (\ref{eq:17}), we calculate the system effective mechanical pressure $P_{m}$, so we have 
\begin{equation}
P_{ef} \rightarrow P_{m}=\frac{2P_{r}+P_{\varphi}}{3}  \label{eq:27}
\end{equation}
hence Eq. (\ref{eq:26}) can be conveniently rewritten as 
\begin{equation}
P_{m}\theta +\frac{\left( P_{m}-P_{\varphi }\right) }{2}\left( \theta
-3\digamma \right) +Q^{\beta }u^{\alpha }\nabla _{\alpha }u_{\beta }+\Omega
^{\alpha \beta }\nabla _{\alpha }u_{\beta }=-u^{\alpha }\nabla _{\alpha }\xi
-\xi \nabla _{\alpha }u^{\alpha }-\nabla _{\alpha }Q^{\alpha }.
\label{eq:28}
\end{equation}

It is important to note that the pressures $P_{r}$ and $P_{\varphi}$ are non-equilibrium pressures. Thus, we define the viscous pressure $\tau$ such that

\begin{equation}
P_{m}=P+\tau,  \label{eq:29}
\end{equation}

where $P$ represents the local equilibrium pressure corresponding to the chosen 4-velocity. Under these conditions, it is hypothesised that $\xi$ matches the energy density of the system in local equilibrium. This assumption ensures that the Euler relation is satisfied,

\begin{equation}
\xi =sT-P+\mu n ,  \label{eq:30}
\end{equation}

where $s$ is the entropy density of the system, $T$ is the temperature of the system, $\mu$ is the relativistic chemical potential, and $n$ is the local equilibrium value of the particle density \cite{is1}. Consequently, the following expression holds,

\begin{eqnarray}
\nabla _{\alpha }\xi &=& T\nabla _{\alpha }s+\mu \nabla _{\alpha }n \label{eq:33a} \\
\nabla _{\alpha }P &=& s\nabla _{\alpha }T+n\nabla _{\alpha }\mu .
\label{eq:33}
\end{eqnarray}

Applying conditions (\ref{eq:33a}), (\ref{eq:33}), the Euler relation (\ref{eq:30}), and the definition (\ref{eq:29}) to equation (\ref{eq:28}), we obtain the following,

\begin{equation}
\Pi \theta +\frac{\left( P+\tau -P_{\varphi} \right) }{2}\left( \theta
-3\digamma \right) +Q^{\beta }u^{\alpha }\nabla _{\alpha }u_{\beta }+\Omega
^{\alpha \beta }\nabla _{\alpha }u_{\beta }+T\nabla _{\alpha }\left(
su^{\alpha }\right) +\nabla _{\alpha }Q^{\alpha }=0,  \label{eq:34}
\end{equation}

where the particle flux conservation $\nabla _{\alpha }\left( nu^{\alpha }\right) = 0$, has been assumed according to \cite{is2}. Furthermore, the term $s u^{\alpha }$ clearly represents a vector associated with the entropy density flux along the worldlines, denoted as $s^{\alpha}$ \cite{eck}. The total entropy $S$ can then be defined by integrating over a spacelike surface $\Sigma$, specified at time $t_0$ as follows,

\begin{equation}
S\left( \Sigma \right) =\int\limits_{\Sigma }s^{\alpha }d^{3}x_{\alpha }.
\label{eq:35}
\end{equation}

The second law of thermodynamics dictates that the total entropy must not decrease over time. Consequently, if $S$ is evaluated over a spacelike surface $\Sigma^{\prime }$, defined at time $t_0+\Delta t$, it must satisfy,

\begin{equation}
S\left( \Sigma ^{\prime }\right) -S\left( \Sigma \right) =\int \nabla
_{\alpha }s^{\alpha }d^{4}x\geq 0,  \label{eq:36}
\end{equation}

where Gauss's theorem has been applied to perform the integration. This condition, together with the fact that $\Sigma^{\prime }$ lies in the future of $\Sigma$, leads to the following inequality,

\begin{equation}
\nabla _{\alpha }s^{\alpha }\geq 0,  \label{eq:37}
\end{equation}

which can be regarded as a sufficient condition for the 
fulfilment of the second law of thermodynamics \cite{his}.

\section{First order thermodynamic framework} \label{four} 

In Eckart's standard theory \cite{eck}, the entropy density flux $s^{\alpha}$ is defined as,

\begin{equation}
s^{\alpha }=su^{\alpha }+Q^{\alpha }T^{-1} \mbox{.}  \label{eq:38}
\end{equation}

Substituting Eq. (\ref{eq:38}) in Eq. (\ref{eq:34}), we obtain the fundamental relation,

\begin{equation}
-\frac{\tau \theta}{T} -\frac{\left( P+\tau -P_{\varphi} \right) }{2T}\left(
\theta -3\digamma \right) -\frac{Q^{\beta }}{T}\left( \frac{\nabla _{\beta }T%
}{T}+u^{\alpha }\nabla _{\alpha }u_{\beta }\right) -\frac{1}{T}\Omega
^{\alpha \beta }\nabla _{\alpha }u_{\beta }=\nabla _{\alpha }s^{\alpha }.
\label{eq:39}
\end{equation}

Finally, and according to \cite{is2} we know that to satisfy condition (\ref{eq:37}), the following must hold,

\begin{eqnarray}
\tau &=&-\zeta \theta \\
Q^{\alpha } &=&-\lambda h^{\alpha \beta }\left( \frac{\nabla _{\beta }T}{T}%
+u^{\gamma }\nabla _{\gamma }u_{\beta }\right) \\
\Omega ^{\alpha \beta } &=&-\eta \left( h^{\alpha \gamma }\nabla _{\gamma
}u^{\beta }+h^{\beta \gamma }\nabla _{\gamma }u^{\alpha }-\frac{2}{3}\theta
h^{\alpha \beta }\right)
\end{eqnarray}

which can be naturally incorporated into our model, allowing us, without loss of generality, to rewrite Eq.~(\ref{eq:39}) as,

\begin{equation}
\frac{\tau ^{2}}{\zeta T}+\frac{\left( P+\tau -P_{\varphi }\right) }{2T\zeta }%
\left( \tau +3\zeta \digamma \right) +\frac{Q^{\alpha }Q_{\alpha }}{\lambda T}%
+\frac{\Omega ^{\alpha \beta }\Omega _{\alpha \beta }}{2\eta T}=\nabla
_{\alpha }S^{\alpha }.  \label{eq:40}
\end{equation}

Thus, it is natural to assume that the following condition must be satisfied,

\begin{equation}
\digamma =\frac{P-P_{\varphi }}{3\zeta }  \label{eq:41}
\end{equation}

so that we ensure the validity of inequality (\ref{eq:37}). 
Consequently, the final form of Eq. (\ref{eq:39}) is given by,

\begin{equation}
\nabla _{\alpha }S^{\alpha }=\frac{\tau ^{2}}{\zeta T}+\frac{\left( \tau +3\zeta \digamma \right) ^{2}}{%
2\zeta T}+\frac{Q^{\alpha }Q_{\alpha }}{\lambda T}+\frac{\Omega ^{\alpha
\beta }\Omega _{\alpha \beta }}{2\eta T} \label{eq:4graS}
\end{equation}

where $\zeta \geq 0$ and $\eta \geq 0$ represent the bulk and shear viscosities, respectively, and $\lambda \geq 0$ denotes the thermal conductivity. These parameters ensure compliance with the second law of thermodynamics and establish a connection between the orthogonal pressures, as described by Eqs. (\ref{eq:41}), (\ref{eq:27}), and (\ref{eq:29}), which can be summarised as follows:
\begin{equation}
P_{\varphi }=P-3\zeta \digamma
\end{equation}

and,

\begin{equation}
P_{r}=P+\frac{3\tau +3\zeta \digamma }{2}
\end{equation}

which indicate how these pressures must relate to an equilibrium pressure for the model to be physically valid. In summary, an anisotropic and irreversible fluid can be expressed in the following form,

\begin{equation}
T^{\alpha \beta }=\xi u^{\alpha }u^{\beta }+P_{r}h^{\alpha \beta }+\left(
P_{\varphi }-P_{r}\right) \widehat{e}_{(3)}^{\alpha }\widehat{e}%
_{(3)}^{\beta }+Q^{\alpha }u^{\beta }+Q^{\beta }u^{\alpha }+\Omega ^{\alpha
\beta },
\end{equation}

which is physically well-defined, provided that,

\begin{eqnarray}
P_{\varphi } &=&P-3\zeta \digamma \label{eq:411}\\
P_{r} &=&P+\frac{3\tau +3\zeta \digamma }{2} \label{eq:42} \\
\digamma &=&\widehat{e}_{(3)}^{\alpha }\widehat{e}_{(3)}^{\beta }\nabla
_{\alpha }u_{\beta } \label{eq:43} \\
\tau &=&-\zeta \theta \label{eq:44} \\
Q^{\alpha } &=&-\lambda h^{\alpha \beta }\left( \nabla _{\beta }T+Tu^{\gamma
}\nabla _{\gamma }u_{\beta }\right) \label{eq:45} \\
\Omega ^{\alpha \beta } &=&-\eta \left( h^{\alpha \gamma }\nabla _{\gamma
}u^{\beta }+h^{\beta \gamma }\nabla _{\gamma }u^{\alpha }-\frac{2}{3}\theta
h^{\alpha \beta }\right).\label{eq:46}
\end{eqnarray}

This set of equations constitutes an extension of the isotropic case and represents one of the contributions to this work. Furthermore, this model provides the foundation to address the isotropic case subsequently, as described in Eckart's thermodynamics. In fact, if $\digamma = 0$, the Eckart thermodynamics for isotropic fluids is recovered.

Natural reasoning suggests that, for models requiring an imperfect fluid, heat fluxes will arise, which must be connected to a fluctuation in the system's mass. 
Consequently, it is natural to assume that the metric must depend on time. A simple example of this case can be a diagonal Schwarzschild-like metric defined by $ g_{00}=-\exp[\nu (r,t)]$, $g_{11}=\exp[\lambda (r,t)]$, $g_{22}=r^{2}$ and $ g_{33}=r^{2} sin^{2}(\theta)$. In this case, 

\begin{equation}
    \digamma = \widehat{e}_{(3)}^{\alpha }\widehat{e}_{(3)}^{\beta }\nabla _{\alpha }u_{\beta }=\frac{u_{0}}{2}\widehat{e}_{(3)}^{3}\widehat{e}_{(3)}^{3}g^{00}\left( \frac{\partial g_{33}}{\partial t}\right) =0,
\end{equation}

which implies that for this type of metric, it is not necessary to consider an anisotropic fluid, as it requires $g_{33}$ to depend on time. This model provides tools to study interior solutions and describe anisotropic models consistent with physical reality.

\section{The second order thermodynamics} \label{five} 

According to  Israel-Stewart formalism for causal thermodynamics for isotropic fluids \cite{is2}, we extend the first-order formulation for anisotropic fluids presented in the previous section to the second-order. Under this approach, the entropy current $s^{\alpha}$ can be expressed by including linear combinations of second-order terms involving dissipative variables. Explicitly, we define,

\begin{equation}
    s^{\alpha }=su^{\alpha }+\frac{Q^{\alpha }}{T}+\frac{C^{\alpha }}{T}, \label{eq:corriente2}
\end{equation}
where $C^{\alpha}$ generalize the typical expression for the entropy current done for the isotropic case, introducing cuadratic terms associated with the anisotropic function $\digamma$ and the new potential thermodynamics $\alpha_{3}$ and $\beta_{3}$, in the form,

\begin{eqnarray*}
   C^{\alpha }&=&-\left( \beta _{0}\tau ^{2}+\beta _{1}Q_{\beta }Q^{\beta }+\beta _{2}\Omega ^{\sigma \nu }\Omega _{\sigma \nu }+\beta _{3}\left( \tau +3\zeta \digamma \right) ^{2}\right) \frac{u^{\alpha }}{2} \\&&+\alpha _{0}\tau Q^{\alpha }+\alpha _{1}\Omega ^{\alpha \beta }Q_{\beta }+\alpha _{3}\left( \tau +3\zeta \digamma \right) Q^{\alpha }.
\end{eqnarray*}

Analogously to the first-order case, we use the second law of thermodynamics, so that from (\ref{eq:corriente2}) we have the following,

\begin{equation}
\nabla _{\alpha }s^{\alpha }=\nabla _{\alpha }   \left(su^{\alpha } \right)+\nabla _{\alpha } \left( {\frac{Q^{\alpha }}{T}}\right)+ +\nabla _{\alpha } \left( {\frac{C^{\alpha }}{T}}\right)\geq 0.
 \end{equation}

Finally, the second-order thermodynamic equations valid for an anisotropic fluid satisfy the following expressions,

\begin{align*}
T \nabla_{\alpha} s^{\alpha} &= 
-\tau \left( \theta + \frac{\beta_0}{2} \tau \theta 
+ u^{\alpha} T \tau \nabla_{\alpha} \left( \frac{\beta_0}{2T} \right) 
+ \beta_0 u^{\alpha} \nabla_{\alpha} \tau 
- \alpha_0 \nabla_{\alpha} Q^{\alpha} 
- \gamma_0 T Q^{\alpha} \nabla_{\alpha} \left( \frac{\alpha_0}{T} \right) \right) \\
&\quad - \frac{(\tau - \tau_{\varphi})}{2} 
\left( (\theta - 3 \digamma) + (\tau - \tau_{\varphi}) 
\left( \beta_3 \theta + 2 T u^{\alpha} \nabla_{\alpha} \left( \frac{\beta_3}{2T} \right) \right) \right. \\
&\quad \left. + 2 \alpha_3 \nabla_{\alpha} Q^{\alpha} 
+ 2 u^{\alpha} \beta_3 \nabla_{\alpha} (\tau - \tau_{\varphi}) 
- 2 \gamma_3 T Q^{\alpha} \nabla_{\alpha} \left( \frac{\alpha_3}{T} \right) \right) \\
&\quad - Q^{\alpha} \left( \frac{\nabla_{\alpha} T}{T} 
+ u^{\lambda} \nabla_{\lambda} u_{\alpha} 
+ \frac{\beta_1}{2} \theta Q_{\alpha} 
+ u^{\lambda} \beta_1 \nabla_{\lambda} Q_{\alpha} 
+ T Q_{\alpha} u^{\lambda} \nabla_{\lambda} \left( \frac{\beta_1}{2T} \right) \right. \\
&\quad \left. - T (1 - \gamma_1) g_{\alpha \gamma} \Omega^{\lambda \gamma} 
\nabla_{\lambda} \left( \frac{\alpha_1}{T} \right) 
- \alpha_0 \nabla_{\alpha} \tau 
- (1 - \gamma_0) T \tau \nabla_{\alpha} \left( \frac{\alpha_0}{T} \right) \right. \\
&\quad \left. - \alpha_1 g_{\alpha \beta} \nabla_{\lambda} \Omega^{\lambda \beta} 
- (1 - \gamma_3) T (\tau - \tau_{\varphi}) 
\nabla_{\alpha} \left( \frac{\alpha_3}{T} \right) 
- \alpha_3 \nabla_{\alpha} (\tau - \tau_{\varphi}) \right) \\
&\quad - \Omega^{\alpha \beta} \left( \nabla_{\alpha} u_{\beta} 
+ \frac{\theta \beta_2}{2} \Omega_{\alpha \beta} 
+ u^{\lambda} \beta_2 \nabla_{\lambda} \Omega_{\alpha \beta} 
+ T \Omega_{\alpha \beta} u^{\lambda} \nabla_{\lambda} \left( \frac{\beta_2}{2T} \right) \right. \\
&\quad \left. - \alpha_1 \nabla_{\alpha} Q_{\beta} 
- T \gamma_1 Q_{\beta} \nabla_{\alpha} \left( \frac{\alpha_1}{T} \right) \right).
\end{align*}

Then, the second-order thermodynamic equations valid for an anisotropic fluid satisfy the following expressions,

\begin{eqnarray}
\tau &=& -\zeta \left( \theta +\phi_{1}\right)  \\
\tau_\varphi
&=& -\zeta \left( 3\digamma-\phi_{2}+ \phi_{1} \right)\\
  \ Q^{\mu }&=&-\lambda h^{\mu \alpha }\chi_{\alpha } \\
  \Omega ^{\alpha \beta }&=&-2\eta \left\langle \omega^{\alpha \beta }\right\rangle,
\end{eqnarray}

where auxiliary functions have been introduced to simplify the expressions, defined explicitly as follows,

\begin{eqnarray*}
\phi_{1} &=& \frac{\beta _{0} \tau}{2} \theta +u^{\alpha }T\tau \nabla _{\alpha }\left( \frac{\beta _{0}}{2T}\right) +\beta _{0}u^{\alpha }\nabla _{\alpha }\tau -\alpha _{0}\nabla _{\alpha }Q^{\alpha }-\gamma _{0}TQ^{\alpha }\nabla _{\alpha }\left( \frac{\alpha _{0}}{T}\right) \\
\phi_{2}
&=&  \left( \tau -\tau _{\varphi }\right) \left( \beta _{3}\theta +Tu^{\alpha }\nabla _{\alpha }\left( \frac{\beta _{3}}{T}\right) \right) +2u^{\alpha }\beta _{3}\nabla _{\alpha }\left( \tau -\tau _{\varphi }\right) \\&&+2\alpha _{3}\nabla _{\alpha }Q^{\alpha }-2\gamma _{3}TQ^{\alpha }\nabla _{\alpha }\left( \frac{\alpha _{3}}{T}\right) \\
\chi _{\alpha }&=&\frac{\nabla _{\alpha }T}{T}+u^{\lambda }\nabla _{\lambda }u_{\alpha }+\frac{\beta _{1}}{2}\theta Q_{\alpha }+u^{\lambda }\beta _{1}\nabla _{\lambda }Q_{\alpha }+TQ_{\alpha }u^{\lambda }\nabla _{\lambda }\left( \frac{\beta _{1}}{2T}\right)\\ 
&& -T\left( 1-\gamma _{1}\right) g_{\alpha \gamma }\Omega ^{\lambda \gamma }\nabla _{\lambda }\left( \frac{\alpha _{1}}{T}\right) -\alpha _{0}\nabla _{\alpha }\tau -\left( 1-\gamma _{0}\right) T\tau \nabla _{\alpha }\left( \frac{\alpha _{0}}{T}\right)\\ && -\alpha _{1}g_{\alpha \beta }\nabla _{\lambda }\Omega ^{\lambda \beta }-\left( 1-\gamma _{3}\right) T\left( \tau -\tau _{\varphi }\right) \nabla _{\alpha }\left( \frac{\alpha _{3}}{T}\right) -\alpha _{3}\nabla _{\alpha }\left( \tau -\tau _{\varphi }\right) \\
\omega ^{\alpha \beta }&=&\nabla _{\alpha }u_{\beta }+\frac{\theta \beta _{2}}{2}\Omega _{\alpha \beta }+u^{\lambda }\beta _{2}\nabla _{\lambda }\Omega _{\alpha \beta }+T\Omega _{\alpha \beta }u^{\lambda }\nabla _{\lambda }\left( \frac{\beta _{2}}{2T}\right) -\alpha _{1}\nabla _{\alpha }Q_{\beta }-T\gamma _{1}Q_{\beta }\nabla _{\alpha }\left( \frac{\alpha _{1}}{T}\right) 
\end{eqnarray*}

where $\left\langle {\,\,}\right\rangle$ denotes the symmetric, purely spatial, and trace-free part of the enclosed tensor, explicitly given by,

\begin{equation}
\left\langle \omega ^{\alpha \beta }\right\rangle =h^{\alpha \rho }h^{\beta \sigma }\left( \frac{\omega _{\sigma \rho }+\omega _{\rho \sigma }}{2}-\frac{1}{3}g^{\mu \nu }\omega _{\mu \nu }h_{\rho \sigma }\right).
\end{equation}

In this way, it is easy to verify that the thermodynamic formulation for a second-order imperfect and anisotropic fluid satisfies the second law of thermodynamics in the same manner as the first-order thermodynamic formulation. In summary, the entropy growth principle holds, ensuring the consistency of the theory at both orders of approximation,

\begin{equation*}
T\nabla _{\alpha }s^{\alpha }=\frac{\tau ^{2}}{\zeta }+\frac{\left( \tau -\tau _{\varphi }\right) ^{2}}{2\zeta }+\frac{1}{\lambda }Q_{\alpha }Q^{\alpha }+\frac{\Omega ^{\alpha \beta }\Omega _{\alpha \beta }}{2\eta }.
\end{equation*}

In conclusion, the Israel-Stewart formulation for the isotropic case is recovered when $\tau_{\varphi} = \tau$.

\section{Conclusions} \label{six}

We developed a thermodynamically consistent formalism for anisotropic, dissipative fluids within a relativistic particle-frame approach, extending both Eckart's first-order theory and Israel-Stewart's second-order theory, explicitly recovering them in the isotropic limit. We incorporate anisotropic pressures, heat fluxes, and viscous stresses while ensuring causality and stability. We also explicitly formulate the equation of the entropy flux and its anisotropic terms consistent with the second law of thermodynamics. In addition, we provided a robust theoretical framework applicable to astrophysical systems influenced by rotation, electromagnetic fields, strong gravitational effects, and any other interaction subject to anisotropy. In particular, numerical simulations of compact stars and accretion flows under varying boundary conditions could be considered. This could unveil new phenomena and deepen our understanding of cosmic systems. Furthermore, incorporating higher-order thermodynamic corrections and coupling with additional fields could provide further insights into the fundamental mechanisms governing these astrophysical systems.


\begin{thebibliography}{99} \baselineskip=1.3\normalbaselineskip


\bibitem{baskey2021} 
L. Baskey, S. Das, and F. Rahaman, "An analytical anisotropic compact stellar model of embedding class I," Mod. Phys. Lett. A \textbf{36}, 2150028 (2021).

\bibitem{becerra2024} 
L. M. Becerra, E. A. Becerra-Vergara, and F. D. Lora-Clavijo, "Realistic anisotropic neutron stars: Pressure effects," Phys. Rev. D \textbf{109}, 043025 (2024).

\bibitem{blandford1982}
R. Blandford and D. Payne, "Hydromagnetic flows from accretion disks and the production of radio jets," Mon. Not. R. Astron. Soc. \textbf{199}, 883–903 (1982).

\bibitem{bocquet1995} 
M. Bocquet, S. Bonazzola, E. Gourgoulhon, and J. Novak, "Rotating neutron star models with a magnetic field," Astron. Astrophys. \textbf{301}, 757 (1995).


\bibitem{bowers1974} 
R. L. Bowers and E. P. T. Liang, "Anisotropic spheres in general relativity," Astrophys. J. \textbf{188}, 657 (1974).

\bibitem{eck} 
C. Eckart, "The Thermodynamics of Irreversible Processes. III. Relativistic Theory of the Simple Fluid," Phys. Rev. \textbf{58}, 919–924 (1940). DOI: 10.1103/PhysRev.58.919

\bibitem{friedman2013} 
J. L. Friedman and N. Stergioulas, \textit{Rotating Relativistic Stars}, Cambridge University Press, Cambridge (2013).

\bibitem{goldreich1982}
P. Goldreich and S. Tremaine, "The dynamics of planetary rings," Annu. Rev. Astron. Astrophys. \textbf{20}, 249–283 (1982).

\bibitem{his} 
W. A. Hiscock and L. Lindblom, "Stability and Causality in Dissipative Relativistic Fluids," Ann. Phys. \textbf{151}, 466–496 (1983). DOI: 10.1016/0003-4916(83)90288-9

\bibitem{is1} 
W. Israel, "Nonstationary Irreversible Thermodynamics: A Causal Relativistic Theory," Ann. Phys. \textbf{100}, 310–331 (1976). DOI: 10.1016/0003-4916(76)90064-6

\bibitem{is2} 
W. Israel and J. M. Stewart, "Transient Relativistic Thermodynamics and Kinetic Theory," Ann. Phys. \textbf{118}, 341–372 (1979). DOI: 10.1016/0003-4916(79)90130-1

\bibitem{krause1980}
F. Krause and K.-H. Rädler, \textit{Mean-Field Magnetohydrodynamics and Dynamo Theory}, Springer, Berlin, Heidelberg (1980).

\bibitem{letelier1980} 
P. S. Letelier, "Anisotropic fluids with two-perfect-fluid components," Phys. Rev. D \textbf{22}, 807 (1980).



\bibitem{roy} 
R. Maartens, "Causal Thermodynamics in Relativity," Lectures given at the Hanno Rund Workshop on Relativity and Thermodynamics (June 1996), arXiv preprint. arXiv:astro-ph/9609119

\bibitem{oppenheimer1939} J. R. Oppenheimer and G. M. Volkoff, "On Massive Neutron Cores," Phys. Rev. \textbf{55}, 374 (1939).






\bibitem{rahmansyah2020} 
A. Rahmansyah, A. Sulaksono, A. B. Wahidin, and A. M. Setiawan, "Anisotropic neutron stars with hyperons: Implication of the recent nuclear matter data and observations of neutron stars," Eur. Phys. J. C \textbf{80}, 769 (2020).


\bibitem{sil} 
R. Silva, J. A. S. Lima, and M. O. Calv\~ao, "On the Thermodynamic Approach to the Cosmological Constant," Gen. Relativ. Gravit. \textbf{34}, 865–882 (2002). DOI: 10.1023/A:1016009126020


\bibitem{DAU}
J. Diaz Polanco, J. Ayala Hoffmann, and M. Ujevic, "The Radiant Massive Magnetic Dipole," Class. Quantum Grav. \textbf{41}, 065017 (2024).


\bibitem{tolman1939} R. C. Tolman, "Static Solutions of Einstein's Field Equations for Spheres of Fluid," Phys. Rev. \textbf{55}, 364 (1939).

\end{thebibliography}
\end{document}